# Serverification of Molecular Modeling Applications: the Rosetta Online Server that Includes Everyone (ROSIE)

Sergey Lyskov*,[1], Fang-Chieh Chou*,[2], Shane Ó Conchúir[5,6], Bryan S. Der[4], Kevin Drew[7], Daisuke Kuroda[1], Jianqing Xu[1], Brian D. Weitzner[1], P. Douglas Renfrew[7], Parin Sripakdeevong[3], Benjamin Borgo[11], James J. Havranek[11], Brian Kuhlman[4], Tanja Kortemme[5,6,12], Richard Bonneau[7,8], Jeffrey J. Gray**,[1,9], Rhiju Das**,[2,10]

[1]Department of Chemical and Biomolecular Engineering, The Johns Hopkins University, Baltimore, Maryland, USA

[2]Department of Biochemistry, Stanford University School of Medicine, Stanford, California, USA

[3]Biophysics Program, Stanford University, Stanford, California, USA

[4]Department of Biochemistry and Biophysics, University of North Carolina at Chapel Hill, USA

[5]California Institute for Quantitative Biomedical Research, University of California, San Francisco, USA

[6]Department of Bioengineering and Therapeutic Sciences, University of California, San Francisco, USA

[7]Department of Biology, Center for Genomics and Systems Biology, New York University, New York, USA

[8]Computer Science Department, Courant Institute of Mathematical Sciences, New York University, New York, USA

[9]Program in Molecular Biophysics, The Johns Hopkins University, Baltimore, Maryland, USA

[10]Department of Physics, Stanford University, Stanford, California, USA

[11]Department of Genetics, Washington University in St. Louis, St. Louis, Missouri, USA

[12]Graduate Group in Biophysics, University of California, San Francisco, USA

* Equal contribution authors
** To whom correspondence should be addressed: *jgray@jhu.edu*, *rhiju@stanford.edu*.

## Abstract

The Rosetta molecular modeling software package provides experimentally tested and rapidly evolving tools for the 3D structure prediction and high-resolution design of proteins, nucleic acids, and a growing number of non-natural polymers. Despite its free availability to academic users and improving documentation, use of Rosetta has largely remained confined to developers and their immediate collaborators due to the code's difficulty of use, the requirement for large computational resources, and the unavailability of servers for most of the Rosetta applications. Here, we present a unified web framework for Rosetta applications called ROSIE (Rosetta Online Server that Includes Everyone). ROSIE provides (a) a common user interface for Rosetta protocols, (b) a stable application programming interface for developers to add additional protocols, (c) a flexible back-end to allow leveraging of computer cluster resources shared by RosettaCommons member institutions, and (d) centralized administration by the RosettaCommons to ensure continuous maintenance. This paper describes the ROSIE server infrastructure, a step-by-step 'serverification' protocol for use by Rosetta developers, and the deployment of the first nine ROSIE applications by six separate developer teams: Docking, RNA *de novo*, ERRASER, Antibody, Sequence Tolerance, Supercharge, Beta peptide design, NCBB design, and VIP redesign. As illustrated by the number and diversity of these applications, ROSIE offers a general and speedy paradigm for serverification of Rosetta applications that incurs negligible cost to developers and lowers barriers to Rosetta use for the broader biological community. ROSIE is available at *http://rosie.rosettacommons.org*.

# Introduction

The Rosetta molecular modeling suite provides tools for a wide range of fundamental questions in structural biology, from the engineering of novel protein enzymes to the prediction of large non-coding RNA structures. The current codebase is rapidly evolving due to the efforts of more than 250 active developers, ease of integrating new functionality into a modular software architecture [1], cross-fertilization between teams working on different systems, and continuing improvements inspired by stringent experimental tests and blind prediction contests [2].

Most of the 50+ applications in the Rosetta package require familiarity with a Unix environment, access to a high-performance computing cluster, and familiarity with tools to visualize and interpret results. A growing number of tutorials, online documentation pages, scripting [3,4] and interactive [3,5,6] interfaces, and introductory papers [1,7] are being written to lower barriers to Rosetta use and development. The most powerful simplification for external users, however, has been in the form of servers. Separate teams of Rosetta developers have created and added functionality to free web interfaces to nine protocols (Table 1) [8-14]. These servers are in high demand from the academic community, with wait times of at least a day in most cases. In some cases, the servers are down. These servers have relied on spare computing resources and administration provided by individual laboratories, rendering them difficult to maintain in the long term. Furthermore, the vast majority of Rosetta applications are not available on servers.

Creating and maintaining web servers – a process we denote 'serverification', in analogy to the term 'gamification' for turning tasks into games – can be complex and laborious. Besides the effort to encode and test the Rosetta protocol, much effort is required to plan database structures, create infrastructure for user interfaces, and other core server tasks. Thus, although the servers in Table 1 have similar components, they all use different application program interfaces (APIs) because they were created in five different labs by different people at different times. The duplicated effort is wasteful. In addition, support and maintenance currently requires sustained effort by each laboratory. In our experience, this post-serverification support can become especially difficult after the researcher who created the server has left the laboratory.

We hypothesized that a common server codebase, a step-by-step serverification protocol, and a virtual machine for testing would lower barriers to server development and thus rapidly accelerate the serverification process. Herein, we present ROSIE, the Rosetta Online Server that Includes Everyone, and demonstrate that it indeed accelerates the rate of serverification. ROSIE presents a common source base that has solved tedious tasks in server implementation and provides developers a simple route to create new servers. The new framework uses a common set of libraries and tools to speed and to simplify the creation of new web interfaces. Additionally, ongoing support of the servers is centralized. Thus, while previous cost-benefit decisions restricted server implementation to broad-use applications such as Robetta [8], RosettaDesign [15], and RosettaDock [10], ROSIE should promote serverification of not only such wide-use applications but also a diverse array of more specific-use and lower-traffic protocols.

Most papers describing new Rosetta functionalities evaluate success through experimental tests of Rosetta predictions of macromolecule structure and behavior. Instead, this paper evaluates success in the ROSIE effort by the rate of creation of novel servers, the extent to which common features are re-used across servers, and the usability of the resulting server (as assessed by number of users and jobs so far). This paper's primary intended audience is the community of current and future Rosetta developers who wish to bring their work into wider use via serverification. Detailed descriptions of each ROSIE application – meant for potential users – are presented in the available online documentation and will also be presented in separate publications elsewhere.

## Results

The following describes the overall ROSIE infrastructure, a detailed 'serverification' protocol that has been used by several developers already, and the successful implementation of shared ROSIE features across nine current applications.

### *A generalized server infrastructure*

Traditionally, Rosetta servers are organized as a front-end web server, a SQL (Structured Query Language) database, a back-end job management daemon and a high-performance computing cluster, all on the same local network (Figure 1). ROSIE implements a more flexible architecture (Figure 2). The server handles multiple protocols feeding the same database, while allowing each lab to customize the appearance of each application and permit uniquely named links from their own web sites and publications.

The following services are implemented:

(a) A generalized database schema that stores a wide range of protocol information. The schema stores input and output files and uses the JSON format (http://www.ietf.org/rfc/rfc4627.txt) for protocol options and other structured data.

(b) Common user interface elements, including a job queue (Figure 3a), user self-registration (Figure 3b), password and account management tasks, and a web-based administrative interface for group, job, and priority management.

(c) For protocol interfaces, user-interface widgets including file uploaders (Figure 3c), Protein Data Bank (PDB) file visualizations (Figure 3d, 3f), and score plot widgets (Figure 3e). Additionally, we have created a library of input validator functions for Python and JavaScript to sanitize (or reject) imperfectly formatted user input and to ensure security.

(d) A layer in the computational back-end to specify what to run (command-line scripts) and how to run them (parallelization scheme and data pipeline). Thus, the job specifications can be used with adapter functions to create scripts for new high-performance computing (HPC) clusters. The current strategy has been designed to consolidate spare computing cycles across multiple resources, and to prepare for the future ubiquitous nature of inexpensive cloud computing resources. At the time of writing, a 320-core cluster is active, with additional plans to make use of a second comparably sized cluster and, in the near future, commercial resources.

The first test application for ROSIE was Rosetta Docking [16], which was ported from a previously existing server. This test provided archetypes of all the functionalities described above (see Fig. 3). At the time of writing, 720 jobs by 133 independent users, totaling 43,870 CPU-hours, have been successfully completed with the ROSIE Docking server.

### *RNA de novo modeling and ERRASER as 'external' test cases*

The basic hypothesis underlying ROSIE was that it would permit rapid serverification of Rosetta applications, even by laboratories at different universities and with different modeling focus than the Johns Hopkins site where the ROSIE code was originally developed. RNA *de novo* modeling [17] and ERRASER [18], both developed by the Stanford Rosetta group, provided first test cases.

RNA de novo modeling, including high resolution refinement (Fragment Assembly of RNA with Full Atom Refinement [17]), was not previously available via a server, but naturally fit into the ROSIE framework. Serverification required a total development time of four weeks (accelerated for later applications; see below). The Stanford team created the application by specifying the inputs, outputs, and Rosetta command-lines for modeling, clustering and simple testing to the main ROSIE administrator (SL). Automated setup of cluster jobs, visualization of model scores vs. RMSD, PyMOL-based rendering of model images, archiving of models, and numerous other features would normally be complex to implement, however, these features were adapted rapidly from existing components from the ROSIE Docking implementation. In addition, features such as text processing and validation of user input before job setup, model post-processing (clustering), interactive display of energy components for each model, and aesthetically pleasing application logos were developed at this time, and later were put into use in other applications. At the time of writing, the ROSIE RNA *de novo* server has been active for approximately one year. The server has completed 170 jobs by 42 separate users, totaling 21,383 CPU-hours. This level of use is notable given that the server was not described in any publication or advertised, aside from two links from the developer's laboratory website at Stanford and from a forum post at the EteRNA project for massively multiplayer online RNA design (http://eterna.stanford.edu).

The third ROSIE application, ERRASER, provided a test case for more rapid and independent implementation by application developers, rather than extensive cross-correspondence between the developers and ROSIE administration. It also was the first example of a Rosetta server being created and published concomitantly with a new Rosetta application, and the first example of a ROSIE server that accepts experimental data (electron density maps). In brief, ERRASER (Enumerative Real-space Refinement ASsisted by Electron density under Rosetta) optimizes the local geometries of RNA crystallographic structures under the constraints of an electron density map and the Rosetta scoring function, and it is designed to be a practically useful tool for RNA crystallographic studies [18]. The Stanford application developers used the previously implemented docking and RNA *de novo* applications as templates and developed the server nearly independently from the ROSIE administrators.

For this third application, a virtual machine image of the server was created to enable local testing of the server by the developer without requiring public deployment on the Web. After the developers were able to run the ERRASER server successfully on their local machine, the ROSIE administrator then integrated the new application to the central server. The overall development time was three weeks (two weeks for the initial deployment by the developers and one week for integrating the application to the central server). The Stanford developers also created a standard protocol for adding new applications to ROSIE (see Methods), which is constantly updated by ROSIE developers.

*Rapid creation of additional server functionalities*

In parallel or after the deployment of the first three applications, a diverse set of seven additional ROSETTA functionalities have been serverified, briefly summarized below.

*β-peptides*

β-peptides are peptides with an additional backbone carbon atom, leading to an extra dihedral angle and an extended length between adjacent side chains. A polymer with a non-biological backbone, structured beta peptides are often called foldamers (see also the NCBB design section below) [19] [20]. Recently, high-resolution structures of multimeric β-peptide bundles have been solved by X-ray crystallography [21,22], which opened up the possibility of performing structure-based rational redesign of the β-peptides [23-26]. A β-peptide redesign protocol was created under the Rosetta framework, and applied to redesign an octameric β-peptide bundle [27]. Briefly, the protocol fixes the backbone of the input model and searches for the lowest-energy combination of the side chains for residues of interest (as specified by the user). For the design of the β-peptide bundle, we also included functionality for symmetric design, in which equivalent residues are forced to have the same side-chain identities and rotamers. All the features mentioned above are available in the ROSIE β-peptide design server. Users input a starting β-peptide structure and specify the residues to be redesigned. One final model with the lowest Rosetta energy is returned as output.

*Adding functionality: NMR chemical shifts in RNA de novo*

NMR chemical shifts have long been recognized as an important source of structural information for functional macromolecules. Backbone chemical shifts are widely used for protein analysis, including determination of protein secondary structures and backbone torsions [28,29] and refinement of three-dimensional models [30-32]. Recently, the integration of non-exchangeable $^1$H chemical shift data with Rosetta RNA *de novo* modeling produced high-resolution RNA structures [33]. Rather than creating a separate ROSIE server, we chose to include the NMR chemical shift guided modeling feature into the ROSIE RNA *de novo* server. This chemical shift guided modeling mode is activated when the user uploads an NMR chemical shift data file during RNA *de novo* job submission. When run in this mode, the RNA structures are first generated by the standard RNA *de novo* method [34,35] and then refined and rescored using a hybrid energy function. The hybrid energy function consists of the standard Rosetta energy function plus a NMR chemical shift pseudo-energy term that is proportional to the sum of squared deviations between experimental and back-calculated chemical shifts. In this mode, the modeling results are the same as in standard RNA *de novo* with three additional data columns reported in the score data table: (1) *rna_chem_shift*, the chemical shift pseudo-energy score, (2) *chem_shift_RMSD*, the root-mean-square-deviation between the experimental and back-calculated chemical shifts, and (3) *num_chem_shift_data*, the total number of experimental chemical shift data points.

*Antibody*

Antibody modeling is important in biological and medical applications, such as antibody design and drug development [36,37]. RosettaAntibody predicts the structure of an antibody variable region given the amino acid sequence of the heavy and light immunoglobulin chains[38]. Originally developed under the Rosetta2 framework [39], RosettaAntibody was available as one of the early public Rosetta web services [40] (Table 1). The protocol identifies the most homologous templates for frameworks of light and heavy chains and each of the complementarity determining region loops (CDR loops). Subsequently, these templates are assembled into a crude model and then CDR-H3 is remodeled with simultaneous $V_L/V_H$ domain orientation optimization and refinement of canonical CDR loops. The implementation of RosettaAntibody available through ROSIE is an expanded and improved version of the previous antibody modeling protocol built on the Rosetta3 platform. The major improvements include the ability to perform loop modeling using the Kinematic Loop Closure (KIC) algorithm [41], a score function restricting CDR-H3 with knowledge-based rules [42], and an updated structural database, which provides better templates for $V_L$, $V_H$, and the canonical CDR loops. The mandatory inputs are the sequences for both the light chain and heavy chain of an antibody. The output includes the coordinates of the antibody $F_V$ models, images of the models, a summary of the homologous templates and scoring information. The output models can be used for the subsequent modeling of an antibody-antigen complex using EnsembleDock [43] or SnugDock [44], both of which are being prepared for future ROSIE implementation.

*Supercharge*

Increasing protein net charge using surface mutations, or supercharging, has many possible uses. Increased net charge can prevent aggregation of partially unfolded states [45,46], thereby improving protein refolding. Improved refolding of protein can increase longevity of protein-based reagents or therapeutics [47] and increase yields when purifying recombinantly-expressed proteins from inclusion bodies [48]. Additionally, highly cationic proteins can undergo nonviral cell entry [49,50] and highly anionic proteins resist kidney filtration for longer retention time in the bloodstream [51]. The potential users of the Rosetta supercharge protocol [52] [53] are experimentalists attempting to enhance the properties of various proteins of interest.

To run the supercharge protocol, the user provides an input PDB file containing the coordinates of the protein structure to be supercharged, and, optionally, an input residue file that can specify positions to leave as wild-type. The input PDB file can be a Rosetta-relaxed crystal structure, a raw crystal structure, an NMR structure, or a homology model. Supercharge operates in fixed backbone mode by default (backbone minimization is optional) and starts by repacking all sidechains. Second, the user specifies the target net charge and either AvNAPSA-mode (Average Neighboring Atoms Per Sidechain Atom mode, implementing a protocol developed by the Liu lab [54]) or Rosetta-mode [52]. As output, the user receives the PDB file of the designed protein, a residue file indicating which residues were allowed to mutate to which amino acid types (this residue file can be subsequently used in other Rosetta design protocols, such as fixed backbone design), and a log file detailing the command-line options, the net charge, the number of mutations, the list of mutated residues, a PyMOL selection text for convenient viewing of mutated residues, and an energy comparison of wild-type residues versus mutated residues for each Rosetta energy term. These Rosetta energies can be used to evaluate Rosetta-mode designs, or they can be used along with the deterministic AvNAPSA-mode as a hybrid approach. For example, AvNAPSA-mode could be used to generate fifteen mutations, and Rosetta energies might flag three of these mutations as energetically unfavorable, resulting in a final list of twelve mutations.

*Sequence Tolerance*

The concept of "tolerated sequence space" describes the set of sequences that are consistent with a protein's structure and function(s). Methods to determine which sequences would be tolerated by proteins, given desired structures and functions, have many uses: designing proteins for new functions or against undesired activities [55], optimizing protein stability [56], anticipating drug resistance mutations [57], or predicting protein interaction specificity [58]. Predicted tolerated sequences can also be used to construct sequence libraries to diversify existing or select for new functions. Such prediction is useful because it is often difficult to accurately identify single successful sequences, especially for functional specifications that are not explicitly modeled in current computational design methods (for example the rate of an enzymatic reaction or the emission maximum of a fluorescent protein [59]).

The Rosetta sequence tolerance protocol [58,60] predicts a set of tolerated sequences for user-defined positions (generally less than ten at a time) in a protein or protein-protein interface, using flexible backbone protein design. The protocol also allows for positions to be mutated before the sequence tolerance simulations (pre-mutation). This is useful, for example, to predict changes in interaction specificity in response to mutations. To emphasize certain functional requirements during sequence design, such as binding to another protein, the protocol allows the weighting of interfaces within and between protein chains.

Two published versions of the protocol are currently implemented in ROSIE. The earlier version [58] was developed originally for PDZ domains (commonly occurring interaction proteins recognizing linear peptide motifs) and has been successfully validated against a large set of phage display data on peptide interaction specificity of natural and engineered PDZ domains [61]. A generalized version [60] was then developed by testing the protocol in several additional systems using a common set of parameters.

Both versions follow the same protocol. First, an ensemble of structures is generated from the starting (or pre-mutated) structure using Monte Carlo simulations involving side chain and backbone moves using the backrub method [62,63]. Next, low-energy sequences are found for each member of the ensemble using the defined interaction weights. Finally, the individual results are combined to create a predicted set of tolerated sequences.

To run the protocol on ROSIE, the user first uploads a protein structure or model in PDB format to the server or enters a PDB ID, using the common ROSIE PDB widget. This widget returns the list of chains and residues from the PDB file so that the user input controls are limited to valid options, reducing the likelihood of accidental errors. The user then selects the participating protein chains (partners) and their internal and pairwise interaction weights. Between one and ten positions can be specified for sequence tolerance prediction ('designed' positions) and up to ten positions can be specified for mutation before prediction. A Boltzmann factor, used to combine predictions from the ensemble, is automatically set to the optimized value determined in the published protocol [58]: this factor may be overridden by the user. Finally, the number of structures to be created for the ensemble is specified. The computation time scales linearly with this value, but higher values allow for more sampling. Because of the limited number of sequences sampled for each structure in the ensemble in the design stage (relative to the number of total sequences possible), we recommend using no more than six designed positions. Future updates to the protocol may allow greater sampling when specifying higher numbers of designed positions.

The output page displays graphics generated using the published analysis scripts [58,60] so that results are presented in a similar fashion to the publications. The user is shown the ranked table of amino acids at the designed positions, individual boxplots per designed position, a sequence motif generated using the WebLogo package [64], and images of ten low-energy structures from the ensemble. ROSIE presents all relevant output files (PDB files, the positional weight matrix, and PNG and PDF versions of the graphics) for download.

*NCBB design*

The NCBB design application designs protein interaction inhibitors with noncanonical backbones (NCBB). NCBBs, also known as peptidomimetics or foldamers, are classes of molecular scaffolds that are a novel strategy to inhibit protein interactions. The NCBB design server application is capable of making inhibitor designs for three different scaffolds, oligooxopiperazines (OOP), hydrogen bond surrogates (HBS) and peptoids. OOPs are molecular scaffolds with a peptide backbone and ethylene bridges between pairs of residues that stabilize the conformation. In this stabilized conformation, OOPs mimic one face of an α-helix ($i$, $i$+4, and $i$+7 residues) and show promise as inhibitors for targets with helices at the interface [65,66]. The HBS scaffold is a peptide where a covalent linker is attached between the 1$^{st}$ and 4$^{th}$ residues. This modification mimics the first hydrogen bond of a helix and stabilizes the peptide into an alpha helix conformation [67] thus improving pharmacokinetic properties. An HBS inhibitor targeting the P300 – Hif1α protein interaction (often poorly regulated in cancer cells) was successful in disrupting the angiogensis pathway in cell based assays [68]. Peptoids are *N*-substituted glycine amino acids, which have known proteolytic resistance and mimic poly-proline type I and type II helices [69]. Many research efforts have shown that peptoids are a valuable avenue for future therapeutics and have shown to be valuable protein interaction inhibitors [70].

In the ROSIE NCBB design application, minor rigid-body perturbations along with backbone specific moves are iterated with the design of residues on the NCBB scaffold. As input, the user submits a Rosetta-formatted PDB file with a target protein and the NCBB scaffold to be designed. Since this application does not do large docking moves, the rigid body conformation of the NCBB scaffold with respect to the target protein needs to be close to the anticipated binding mode in order to achieve a successful design. The user also can specify which residues to design on the NCBB scaffold as well as how many design cycles and perturbations per cycle the application will perform. The application selects a single final design that is chosen from a filtered set of all decoys produced (top 5% of total_score) and sorted by binding energy (REPACK_ENERGY_DIFF). Users can download the score file with additional scoring information and all decoys produced can be found in the decoys directory.

*RosettaVIP*

One of the first observations of early crystal structures is that the hydrophobic cores of proteins are well-packed [71]. Cavities in the cores of proteins are associated with loss of stability and conformational specificity. Computational filters, such as RosettaHoles, can identify packing defects in computationally designed structural models [72]. Typically, models with defects are discarded. The RosettaVIP protocol identifies mutations that are predicted to improve hydrophobic packing in the cores of proteins, and therefore to generate mutants with enhanced stability [73]. The application of this protocol has been shown to "rescue" protein designs with packing defects. The protocol has also proven itself capable of suggesting mutations that can improve the stability of wild-type proteins.

The RosettaVIP protocol is iterative. Each iteration either suggests a mutation that is predicted to improve the packing of the protein core, or terminates the protocol if no such mutation can be identified. The output of one iteration (with the new mutation and a relaxed structural model) is the input for the next iteration. The protocol takes as initial input a structural model in PDB format. The user selects the total number of iterations (i.e., the maximum number of mutations) to try; the protocol may be configured to run iterations until no further mutations are found. Because the protocol relies on a stochastic version of the RosettaHoles algorithm to identify protein cavities, it is sometimes beneficial to retry an iteration that fails to find a mutation in the hopes that a second attempt will succeed. The user specifies the maximum number of failed attempts at each iteration before the protocol is terminated. Finally, the user may specify a list of residues that should be excluded from mutation, if some prior activity of the original protein is to be preserved. The output of the protocol is a structural model in PDB format that incorporates all suggested mutations and all structural relaxation performed during the course of the protocol. Optionally, intermediate structural models at the end of each successful iteration may be generated as output.

## Discussion

We have developed a core web server infrastructure called ROSIE, the Rosetta Online Server that Includes Everyone, to lower barriers to Rosetta application deployment as servers. ROSIE was presented at the 2012 Rosetta Developers Conference (August, 2012) to illustrate its speed in implementing the RNA de novo application and to identify new protocols needing web distribution. After this conference, the first ROSIE applications RosettaDock and RNA-Denovo [14] were joined by ERRASER [15], RNA *de novo* with chemical shifts, Antibody, Sequence Tolerance, Supercharge, Beta peptide design, NCBB design, and VIP (Table 2). Additional servers in preparation are listed in Table 2. The number of these applications, created on a timescale of a few months, is similar to the number of applications previously implemented in several years of Rosetta development, supporting our hypothesis that the ROSIE framework would accelerate serverification. These ROSIE protocols are now online, free of charge for academic users, at *http://rosie.rosettacommons.org*. From October 2012 to the time of writing (January 2013), more than 550 users have registered and more than 1319 jobs have been completed. These jobs reflect more than 65,000 CPU-hours of modeling computational leveraged by the general biological and modeling communities.

The design philosophy of ROSIE was meant not only to accelerate serverification, but to promote maximal sharing of data and collaboration. This emphasis on sharing follows from the premise of the RosettaCommons initiative – a shared source code and a continuous open flow of scientific information, mostly funded by public research funds. Six features of the present manuscript derived from this largely open philosophy. First, this manuscript is being submitted to an open access journal that does not force privacy restrictions on described servers, unlike other journals. Second, all ROSIE input forms and documentation are open for anyone to view without registration. (Users are encouraged to register an account to track their jobs, but this is not required. Those who register receive email notifications for job submission, start, and finish, and a link to the job status page.) Third, documentation for how to use the server is a prerequisite for deployment, promoting the writing of user-friendly explanations of applications by developers. Fourth, by default, all input and output data are shared publicly on the job queue, although for those concerned about privacy there is an option to hide job output. Fifth, as a further incentive for openness, jobs that are available for open access are given priority access to the ROSIE computing resources. Sixth, the job output can be easily shared with others though email or via social networking widgets including Facebook and Google+.

Our long-term goal is to provide free web versions of all core Rosetta protocols. In the near future, Rosetta developers are planning to use ROSIE to serverify a wide range of applications, covering the full spectrum of available functionalities: small molecule docking [20], multi-state design [21], flexible peptide docking [9] [10], enzyme design [22], $pK_a$ prediction [74], and scaffold grafting [23]. With the server tools already implemented and detailed documentation available for new developers, the web server creation process and ROSIE maintenance has become streamlined, although we anticipate two areas of improvement. First, although presently the ROSIE computational resources are not over-burdened, additional applications and users may eventually strain the system. To counteract this, ROSIE draws from a large pool of RosettaCommons laboratories and will also gain funds from commercial users. Plans are being made to incorporate larger public clusters into the ROSIE backend and to enable the collection of fees from commercial users to support growing infrastructure. Second, complex workflows requiring multiple stages of calculations whose results depend on previous calculations (e.g. stepwise assembly of RNA motifs [75], RasRec for refining large proteins with NMR data [76], or homology modeling including identification of templates [8]) cannot yet be implemented. In these cases, subcomponents of the workflows can be serverified to permit potential users to preview steps of the more complex workflows. Finally, we note that emerging scripting systems [3,4] and interactive interfaces to Rosetta [5,6] are lowering barriers to Rosetta applications through laptops. ROSIE may offer a useful backend to these services as their users require more computational power.

**Acknowledgements**
We are grateful to P. Cordero for help in designing the RNA de novo server.

**Funding**
We acknowledge financial support from the US National Institutes of Health (R01-GM073151 to B.K., J.J.G. and S.L.; R01-GM078221 to J.J.G., R21-GM102716 to R.D., R00-RR024107 to J.J.H., U54CA143907-01 and PN2 EY016586-06 to R.B. and P.D.R. and K.D., T32 GM 88118-2 to K.D.), a Burroughs-Wellcome Career Award at Scientific Interface (R.D.), Governmental Scholarship for Study Abroad of Taiwan and Howard Hughes Medical and Institute International Student Research Fellowship (F.-C.C.), the DARPA Antibody Technology Program (HR-0011-10-1-0052) for J.X. and D.K. S.Ó.C and T.K. were supported by grants from the National

Science Foundation (MCB-CAREER 0744541, EF-0849400, EEC-0540879) and the UC Lab Research Program. R.B., P.D.R. and K.D. were supported by US National Science Foundation (CHE-1151554 and NSF IOS-1126971 to R.B. and P.D.R. and K.D.)

**Methods**

*ROSIE server*
The ROSIE server was implemented with PostgreSQL as a database, using the TurboGears web server framework. Dynamic web controls and widgets were implemented in jQuery, jQuery-UI, jqGrid and FlotCharts libraries. The code for the ROSIE server is under version control, and available to all Rosetta developers, through the same repository that hosts the Rosetta codebase: **svn.rosettacommons.org/trac/browser/trunk/rosie**. The applications are freely available on the World wide Web to the public at **http://rosie.rosettacommons.org**.

*Protocol for 'serverification' of a new Rosetta application*
The following summarizes the steps required for a developer to turn an existing Rosetta application into a ROSIE server. It is a snapshot of the protocol at the time of writing. A continuously updated version of this protocol is being made available at http://goo.gl/Sh7oB. Importantly, the protocol has been written by new ROSIE developers and so captures the perspective required to promote faster first development cycles for other new engineers. ROSIE development tools and source code are available to registered developers through RosettaCommons.

I. Install a local ROSIE test server

Download the VM (http://graylab.jhu.edu/ROSIE) and open it with VirtualBox (http://www.virtualbox.org). Before you start, you may want to do a `svn update` in `~/rosie` and `~/R/trunk/rosetta`, and rebuild the Rosetta trunk, since they may be out of date.

1. Modify the file `rosie/rosie.front/development.ini`. Find the line `host = 192.168.0.64` and comment it out. Enable the line `host = 127.0.0.1`.

2. To run the server: Open two terminals. In one of them, cd into `rosie/rosie.back` and execute `./run_rosie-daemon.sh`. In the other terminal, cd into `rosie/rosie.front` and execute `../run-rosie-server.sh`.

3. Open `localhost:8080` in your browser. Login as admin (password: managepass).

II. Add a new application XXX (Tip: check other released apps to see how to format the files):

1. Create your application in `rosie.back/protocols/XXX`. You need at least two files: `submit.py` and `analyze.py`. See "`rna_denovo`" for example files.

2. For machine-dependent files, edit `rosie.back/data.template/XXX`. Edit `rosie.back/rosie-daemon.ini.template`, add useful shorthands and add the app into the protocol line. Copy the corresponding files to `rosie.back/data/XXX` and `rosie.back/rosie-daemon.ini` so the VM server can read the files.
3. Add the corresponding controller in `rosie.front/rosie/controllers/XXX.py`. See `rna_denovo.py` as an example.
4. Add your controller into `controllers/root.py`. In `root.py`, search for `'rna_denovo'`. Add the two corresponding lines for your application.
5. During the creation of the controller files, you may want to make some validation checks for the input format. They are in `rosie.front/rosie/lib/validators`. You need to create your own validation tests.
6. Create your page in `rosie.front/rosie/templates/XXX/`. You need at least 3 pages: `index.html`, `submit.html`, and `viewjob.html`. See `rna_denovo` for example.
7. Link your application to the main page in `template/index.html`.
8. You may want an icon. Put a png file of ~ 1024*1024 into `rosie/public/image/XXX_icon.png`, and link it to the pages.
9. For documentation, create pages in `template/documentations`. Also you need to edit `controllers/documentation.py` to let the server know where it is. Then link your documentation to `documentation/index.html` and in the other pages of your application.
10. Edit `rosie.front/rosie/websetup/bootstrap.py` and add the name of the new app.
11. Go to `rosie.front/`. Run `'source ~/prefix/TurboGears-2.2/bin/activate'` then `'python update_protocol_schema.py'` to update the database.
12. Test the new application in the browser of the VM to make sure it runs fine.
13. Create a new file `rosie/doc/XXX.txt`, put a short description of protocol input, output, and command line flags. Also add an example job, with input files and a simple readme, into `rosie/examples/validation_tests`.
14. Commit the changes (use `'svn commit --username XXXX'` to specify the user name of the commit). Inform the ROSIE administrators for integration into the central server.

**Table 1**: Public Rosetta servers available prior to current work, in chronological order of development.

| Application | Server | Year | Jobs | Developer | Status | References |
|---|---|---|---|---|---|---|
| *Ab initio, fragments, alascan* | Robetta.org | 2004 | 34000 | Baker@UW | 7-day queue | [8] |
| **Design** | rosettadesign.med.unc.edu | 2006 | 17022 | Kuhlman@UNC | 1 day queue | [15] |
| **Antibody** | antibody.graylab.jhu.edu | 2007 | 2437 | Gray@JHU | Offline | [9] |
| **Docking** | rosettadock.graylab.jhu.edu | 2007 | 10000 | Gray@JHU | 3-7 day queue | [10] |
| **FunHunt** | funhunt.furmanlab.cs.huji.ac.il | 2008 | 173 | Furman@HebrewU | 1 day queue | [11] [12] |
| **FlexPepDock** | flexpepdock.furmanlab.cs.huji.ac.il | 2010 | 5000 | Furman@HebrewU | 1 day queue | [13] [14] |
| **Backrub** | kortemmelab.ucsf.edu/backrub | 2010 | 4,300 | Kortemme@UCSF | 1 day queue | [63] [77] [78] [58] [60] |

**Table 2**: Applications made available via the Rosetta Online Server Including Everyone (ROSIE), in chronological order of development.

| ROSIE Application | Year | Jobs | Developer | Status | References |
|---|---|---|---|---|---|
| **Docking** | 2011 | 752 | Gray@JHU | Public | [10] |
| **RNA Denovo (with NMR chemical shifts)** | 2012 | 177 | Das@Stanford | Public | [33,79] |
| **Erraser** | 2012 | 8 | Das@Stanford | Public | [18] |
| **Beta Peptide Design** | 2012 | 5 | Das@Stanford | Public | [19] |
| **Antibody** | 2013 | - | Gray@JHU | Testing | [38] |
| **Sequence Tolerance** | 2013 | - | Kortemme@UCSF | Testing | [58] [60] |
| **Supercharge** | 2013 | 46 | Kuhlman@UNC | Public | [52,80] |
| **NCBB design** | 2013 | - | Bonneau@NYU | Testing | [81] |
| **VIP** | 2013 | - | Havranek@WUSTL | Testing | [73] |
| **pKa** | 2013 | | Gray@JHU | In preparation | [82] |
| **EnsembleDock** | 2013 | | Gray@JHU | In preparation | [83] |
| **SnugDock** | 2013 | | Gray@JHU | In preparation | [44] |
| **Ligand docking** | 2013 | | Meiler@Vanderbilt | In preparation | [84] |

**Figure 1.** Schematic of a typical server for molecular modeling.

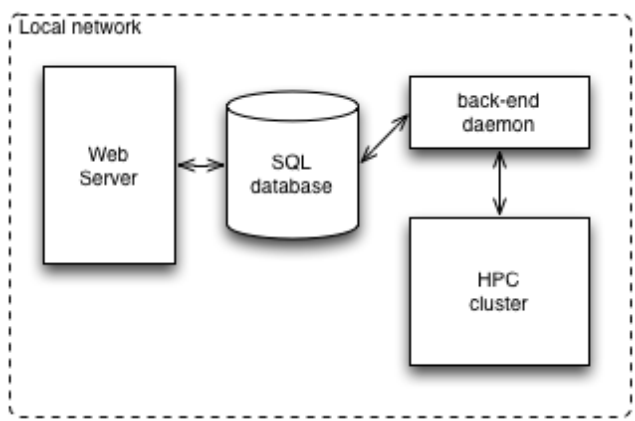

**Figure 2.** Schematic of ROSIE (Rosetta Online Server that Includes Everyone), which permits a number of front-ends for job submission by users, a number of servers (stored in a unified database), and a number of backends to allow expansion of computational resources.

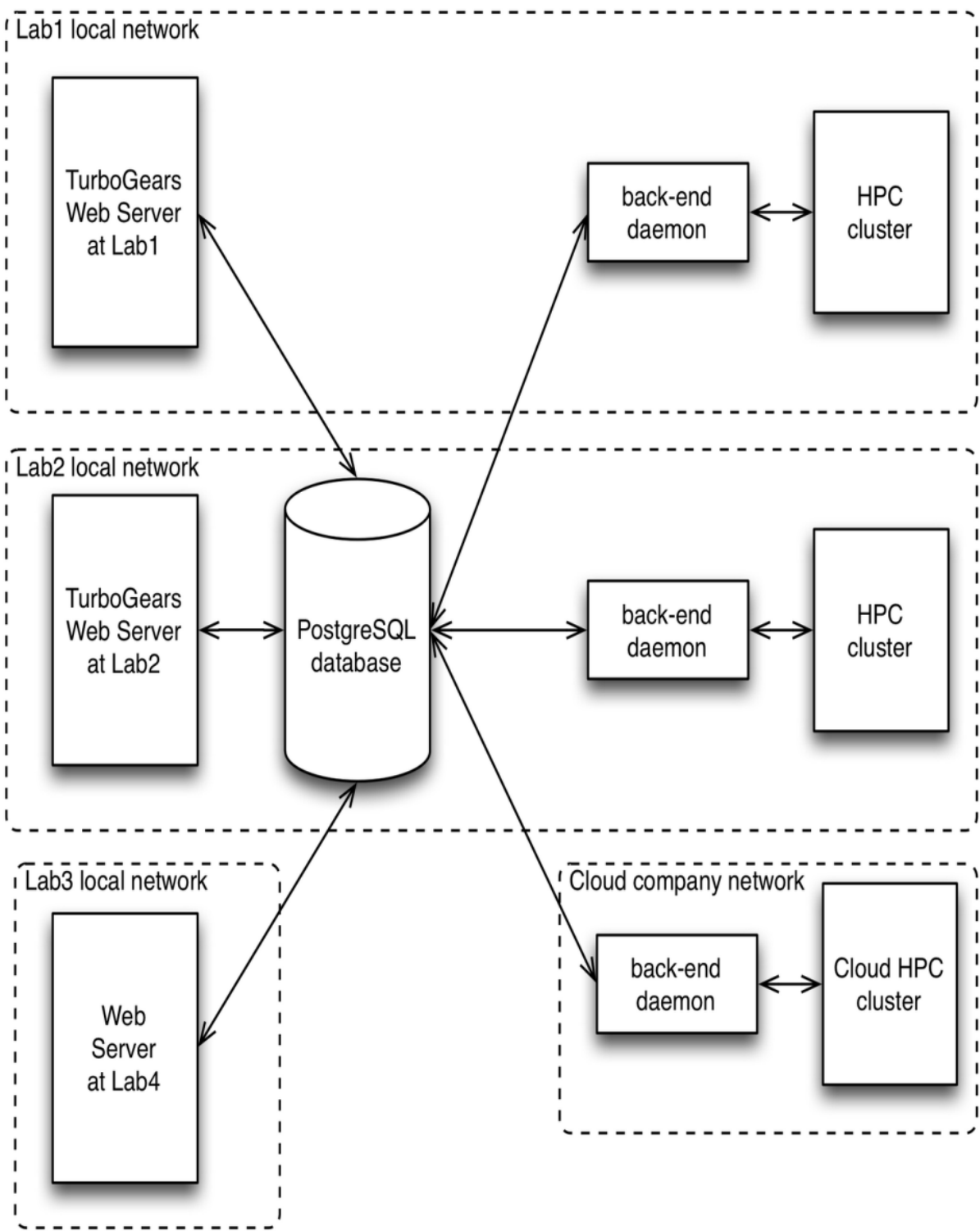

**Figure 3.** Examples of re-usable features and widgets shared across ROSIE servers. **(a)** Global job queue page, which can be filtered by specific application (e.g., docking). **(b)** Self-registration (not required). **(c)** Coordinate file uploader using Protein Databank format, **(d)** Automatic visualization of uploaded coordinate file, **(e)** Score vs. root mean squared deviation plotting widget, **(f)** Automatic rendering of final models, which can be customized by developer for specific applications (in this case, RNA *de novo* modeling).

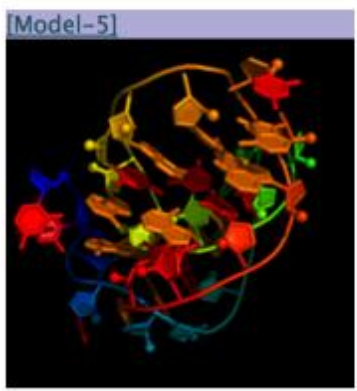
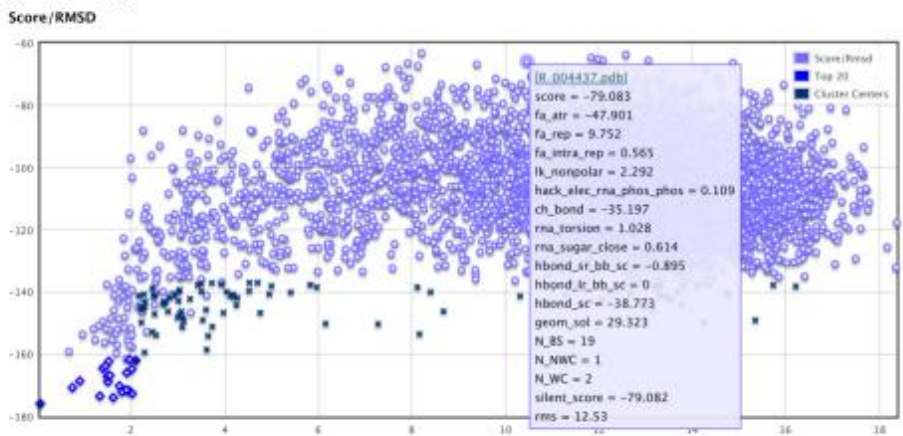